\documentclass[twocolumn,amsmath,amssymb,superscriptaddress]{revtex4}

\usepackage{graphicx}
\usepackage{dcolumn}
\usepackage{bm}

\graphicspath{%
    {figs/}
}
\usepackage{color}
\usepackage{times}

\begin{document}
\preprint{APS/123-QED}

\title{Spatially-Resolved Temperature Measurements of Electrically-Heated Carbon Nanotubes}

\author{Vikram V. Deshpande}
\altaffiliation[Present address: ]{Department of Physics, Columbia University, New York, NY 10027}
\email{vdesh@phys.columbia.edu}
\affiliation{Applied Physics, California Institute of Technology, Pasadena, CA 91125}
\author{Scott Hsieh}
\affiliation{Applied Physics, California Institute of Technology, Pasadena, CA 91125}
\author{Adam W. Bushmaker}
\affiliation{Department of Electrical Engineering, University of Southern California, Los Angeles, CA 90089}
\author{Marc Bockrath}
\affiliation{Applied Physics, California Institute of Technology, Pasadena, CA 91125}
\author{Stephen B. Cronin}
\email{scronin@usc.edu}
\affiliation{Department of Electrical Engineering, University of Southern California, Los Angeles, CA 90089}
\date{\today}

\begin{abstract}
Spatially-resolved Raman spectra of individual pristine suspended carbon nanotubes are observed under electrical heating. The Raman G$_+$ and G$_-$ bands show unequal temperature profiles. The preferential heating is more pronounced in short nanotubes (2 $\mu$m) than in long nanotubes (5 $\mu$m). These results are understood in terms of the decay and thermalization of non-equilibrium phonons, revealing the mechanism of thermal transport in these devices. The measurements also enable a direct estimate of thermal contact resistances and the spatial variation of thermal conductivity.
\end{abstract}
\pacs{Valid PACS appear here}
\maketitle
The temperature of a macroscopic solid is manifest in the energy of its lattice vibrations, or phonons. In nanostructures, this definition of temperature can break down if the phonon population is driven out of equilibrium by an electrical current. When that is the case, some phonon modes can have higher effective temperatures than the rest of the lattice depending on their coupling with the current. In the past, several interesting observations have been made in electrically-heated carbon nanotubes such as current saturation \cite{yaoPRL2000}, $\Gamma$- and K-point optical phonon scattering \cite{parkjyNL2004,javeyPRL2004}, ballistic phonon propagation \cite{chiuPRL2005} and electrical breakdown \cite{collinsPRL2001}. In freely suspended carbon nanotubes, where phonon relaxation is hindered due to the absence of a substrate acting as a thermal sink, striking negative differential conductance (NDC) at high electric fields has been observed \cite{popPRL2005} and explained using non-equilibrium or ``hot'' phonons \cite{popPRL2005,lazzeriPRB2006}.

These high-field properties are particularly relevant for applications of carbon nanotubes as field-effect transistors and interconnects towards the miniaturization of electronics. Additional insight into these properties can be gained through simultaneous optical and transport studies. In particular, because nanotubes are one-dimensional structures with a huge aspect-ratio, these phenomena can vary spatially, necessitating a local probe of temperature in order to fully understand thermal transport in nanotubes. Previously, scanning force microscopy \cite{kimPBCM2002} and local melting of nano-particles \cite{begtrupPRL2007} have been used to extract local temperatures of multi-walled nanotubes under high bias. However, these contact-based techniques involve temperature drops at the measurement interface, hampering their ability to accurately measure temperature.

Raman spectroscopy is a powerful, non-contact method of probing phonons in nanotubes \cite{dresselhaus_raman_2002}. This technique enables one to probe the $\Gamma$-point longitudinal optical (LO), transverse optical (TO) and the radial breathing mode (RBM) phonons, among others. Recently, hot phonons in nanotubes have been directly observed using Raman spectroscopy in conjunction with electrical transport \cite{adam_w._bushmaker_direct_2007,oron-carlPRL2008}, using the temperature response of the Raman G band in nanotubes \cite{huong_temperature-dependent_1995,huang_temperature_1998,raravikar_temperature_2002}. Thus, spatial investigation of electrical heating using Raman spectroscopy forms the motivation for our experiment.

\begin{figure}[t] 
  \centering
  \includegraphics[width=2.5in,keepaspectratio]{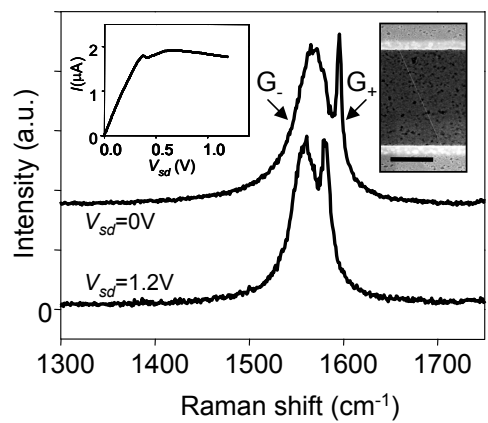}
  \caption{Raman spectra of the center of device D1 at two bias voltages. Left inset: {\it I-V} characteristics of D1. Right inset: SEM viewgraph of a typical device. Scale bar corresponds to 2 $\mu$m.}
  \label{fig:fig1}
\end{figure}

Individual single-walled carbon nanotube devices were grown on top of Pt leads, as reported in detail previously \cite{wigner}. Devices used in this work were grown using ethanol as the carbon feedstock, which has been shown to yield long and low-disorder nanotubes \cite{huangJPCB2006}. Figure 1 (right inset) shows a scanning electron microscope image of a 5 $\mu$m long nanotube device. Raman spectra were measured in a Renishaw InVia spectrometer with Spectra-Physics 532 nm solid state and 785 nm Ti sapphire lasers. Typical integration times were 60-120 seconds. An Ithaco current preamplifier was used to measure the current through the nanotube. All measurements were performed at room temperature, and in an argon environment to prevent burn-out of devices at high bias voltages. Figure 1 (left inset) shows the current-voltage ({\it I-V}) characteristics of a typical quasi-metallic device D1 (5 $\mu$m long). Note the NDC, characteristic of suspended devices, above $V_{sd}\sim0.6$ V in D1.

Figure 1 shows Raman spectra taken at the center of D1 using a 532nm laser at two bias voltages, $V_{sd}$=0 V and 1.2 V. Note the narrow G$_+$ and broad G$_-$ peaks that are characteristic of the TO and LO phonon modes, respectively, of metallic nanotubes. The two peaks downshift unequally in energy on application of a 1.2 V bias voltage. A low enough laser power was used such that the laser itself did not cause any downshift or heating. Note that the defect-induced D band peak in the Raman spectrum, occurring in typical nanotubes around 1350 cm$^{-1}$, is absent in most of our devices, as it is in D1. This attests to the low-defect nature of our devices.

\begin{figure}[t] 
  \centering
\includegraphics[width=3.25in,keepaspectratio]{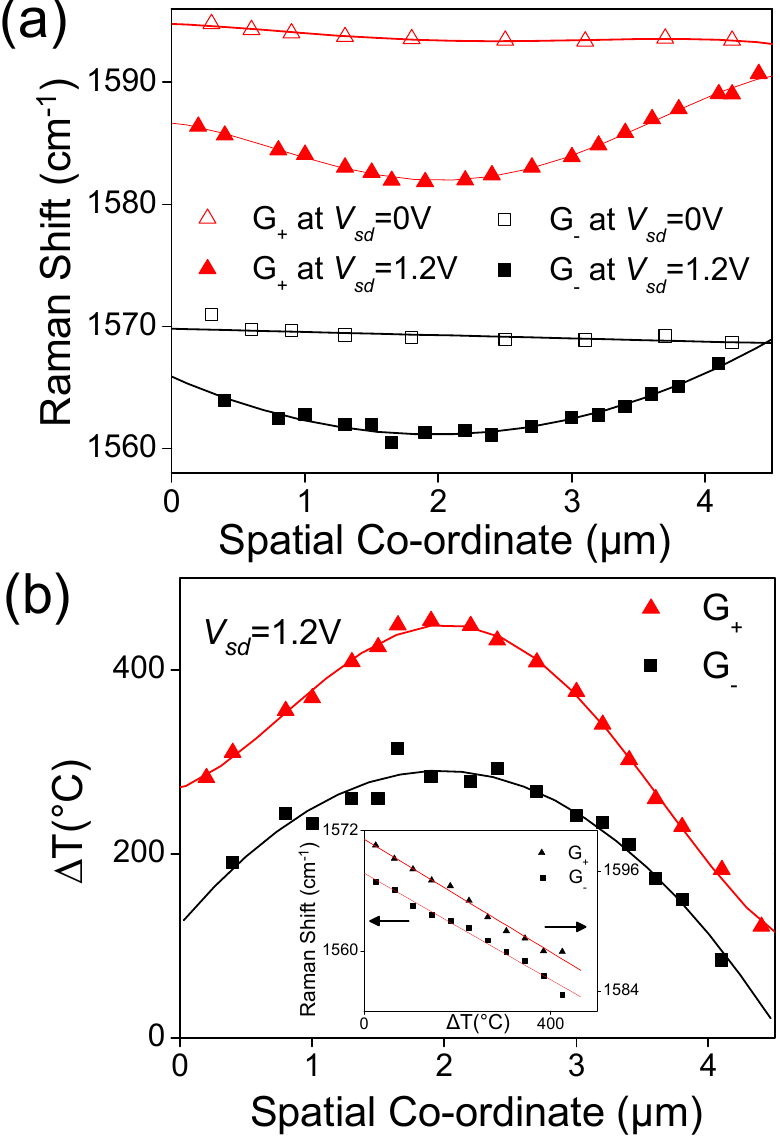}
  \caption{(Color online) (a) Spatial profile of Raman shifts of G${_+}$ and G${_-}$ bands for device D1 at 0 V and 1.2 V bias. (b) Temperature profiles of G$_+$ and G$_-$ bands. Inset: G$_+$ and G$_-$ shifts for D1 in a temperature-controlled stage.}
  \label{fig:fig2}
\end{figure}

We take spatially resolved Raman spectra of our devices at different bias voltages. The 532 nm laser with a diffraction-limited spot-size of 360 nm, affords 10-20 data points along the length of D1. Figure 2(a) shows the Raman shifts of the G$_+$ and G$_-$ bands, along the length of the nanotube, for $V_{sd}$=0 V and 1.2 V. At $V_{sd}$=1.2 V, the bands downshift significantly and develop a spatial profile. The G$_+$ profile fits to a fourth order polynomial, while a parabola suffices for the G$_-$ profile. The Raman shifts observed at high bias are subtracted from the reference ($V_{sd}$=0 V) at every spatial point. This nullifies any contributions from the local nanotube environment to the downshift profile.

Previously, we showed that the downshift of the G$_+$ and G$_-$ bands could be interpreted in terms of increased phonon populations \cite{adam_w._bushmaker_direct_2007} and hence effective temperatures. This interpretation was in agreement with the Stokes/anti-Stokes intensity ratio observed in that work. In this work, we calibrated the G band downshifts with temperature in a temperature-controlled stage. This calibration data is shown in the inset of Fig. 2(b) for device D1. The downshift is linear in temperature, with almost equal slopes for the two bands. The G band temperature is obtained at each point along the nanotube by dividing the voltage-induced change in the Raman shift by the slope of the calibration line. The resulting temperature profiles for the two bands are shown in Fig. 2(b).

This result constitutes the first observation of a spatial temperature profile of a single-walled nanotube under Joule heating. Figure 2(b) also shows that the temperatures at the ends of the nanotube are higher than room temperature and highly asymmetric, indicative of asymmetric thermal contact resistances. Of ten 5 $\mu$m devices studied, 3 devices exhibited G$_+$ and G$_-$ bands that could be resolved separately with bias voltage. All three devices showed the same qualitative behavior described above. Note that the G band Stokes/anti-Stokes intensity ratio was not resolvable with bias for any of these three devices.

\begin{figure}[b] 
  \centering
  \includegraphics[width=3.25in,keepaspectratio]{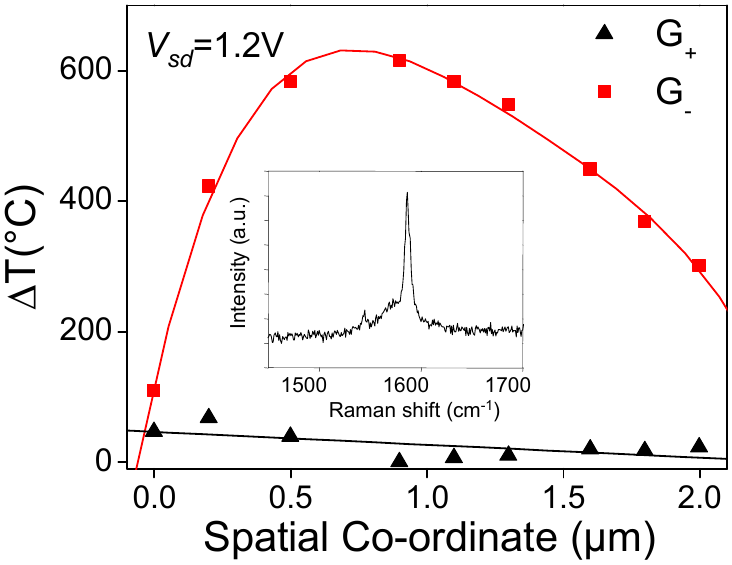}
  \caption{(Color online) Spatial temperature profiles of G$_+$ and G$_-$ bands. Inset: Raman spectrum at the center of D2.}
  \label{fig:fig3}
\end{figure}

In an earlier work \cite{adam_w._bushmaker_direct_2007}, we reported that only one of G$_+$ or G$_-$ bands downshifted with bias. This observation was most prominent in shorter devices ($\leq2 \mu$m). Figure 3(inset) shows the Raman spectrum for such a 2 $\mu$m device, D2. On application of $V_{sd}$=1.2 V to this device, the broad G$_-$ band (LO) downshifted by 22 cm$^{-1}$, while the other bands (G$+$ and another mode at $\sim$1545 cm$^{-1}$) remained unchanged within 1 cm$^{-1}$, similar to results reported in \cite{adam_w._bushmaker_direct_2007}. Figure 3 shows the (converted) effective temperature of the G$_+$ and G$_-$ bands at $V_{sd}$=1.2 V, along the length of D2. Thus, while the G$_-$ band shows a non-uniform temperature profile, the G$_+$ and other band at 1545 cm$^{-1}$ (not shown, to maintain clarity) show no modulation in space (within the error of the measurement).

We now interpret these results. In thermal equilibrium, both G$_+$ and G$_-$ downshift equally, as observed in the temperature-controlled stage data of Fig. 2(b)-inset. Even in the case of thermal non-equilibrium, each phonon mode is likely to be at least at hot as the lattice. The lower temperature profiles shown in Figs. 2(b) and 3 are, thus, upper bounds on the lattice temperature profile. In Fig. 3, this lower profile coincides with room temperature; therefore, the observation that the G$_+$ mode of D2 does not modulate with bias or space indicates that the lattice remains at room temperature. In contrast, the G$_-$ mode exists at an elevated non-equilibrium effective temperature, as a direct result of generation by high-energy electrons. In long nanotube devices, these hot phonon modes transfer energy to the pool of thermalized phonons through inter-phonon scattering processes, heating up the rest of the lattice.

\begin{table}[t]
\caption{\label{tab:table1}Bias response of the G band (which of G$_+$ and G$_-$ respond to bias voltage) versus approximate length $L$ and G$_+$/G$_-$ intensity ratio for several devices.}
\begin{ruledtabular}
\begin{tabular}{lccc}
 No. &$L$ ($\mu$m) &Bias response &Intensity ratio\\
\hline
1 & 0.5 & G$_+$ & 0.34\\
2 & 0.5 & G$_+$ & 0.07\\
3 & 1 & G$_+$ & 0.45\\
4&2&G$_-$&0.11\\
5&2&G$_+$&0.42\\
6 (D2)&2&G$_-$&1.53\\
7&2&Both&0.04\\
8&2&Both&0.09\\
9&5&Both&0.04\\
10 (D1)&5&Both&0.12\\
11 (D3)&5&Both&3.99\\
\end{tabular}
\end{ruledtabular}
\end{table}

We note that while the G$_-$ bands of D1 and D2 are broad, indicative of strong electron-phonon coupling present in metallic nanotubes \cite{wu_variable_2007}, their intensities differ strongly. This is likely the effect of differing chiralities for D1 and D2 \cite{saito_chirality-dependent_2001}. To study whether chirality plays an important role in our observations, we tabulate in Table \ref{tab:table1}, the bias response of the G band with length and G$_+$/G$_-$ intensity ratio for several devices. Table \ref{tab:table1} confirms that length, rather than chirality, is the primary difference between equilibrium and non-equilibrium behavior.

For a quantitative explanation, we first model the hot phonon profile for the simpler case of no lattice heating (Fig. 3). In this case, thermal transport is not governed by Fourier's law (diffusive heat transport), since the thermalized phonons are not heated. Instead, one needs to understand the decay process of the hot phonons. Consider a nanotube of length $L$ under uniform heating by an electric current $I$. Let $g(I)$ be the hot phonon generation rate per unit length, $d$ be the phonon decay length  and $\tau$ be the decay time. Hence, a phonon generated at point $x$ has a probability of reaching point $x_0$ given by $exp(-|x-x_0|/d)$, which is a general characteristic of decay processes. Considering phonons arriving from both directions and a decay rate of $n/\tau$, where $n$ is the phonon population at any point, the continuity equation gives:
\begin{eqnarray}
\frac{dn}{dt}=\int_0^{x_0} g(I) e^{-\frac{x_0-x}{d}} dx+\int_{x_0}^L g(I) e^{-\frac{x-x_0}{d}} dx - \frac{n}{\tau} = 0
\end{eqnarray}
Plotting the solution for $L$=2 $\mu$m, one obtains a good match to the temperature profile of G$_-$ in Fig. 3 using a decay length $d\sim1 \mu$m.

We also need to understand the mechanism of power dissipation. In device D2, at $V_{sd}$=1.2 V and peak $I\sim$5 $\mu$A, the power input to the nanotube is 6 $\mu$W, carried away mainly by thermal conduction along the nanotube. The electronic contribution to thermal conduction is known to be a small fraction (1/5th or less) of the equilibrium lattice contribution \cite{yamamotoPRL2004}. An upper bound to the  lattice thermal conductance, $G_{th}$, of nanotubes in quasi-equilibrium is given by the result for ballistic phonons \cite{footnote_ballistic}. At room temperature, $G_{th}\sim8r_t$ W/K \cite{mingoPRL2005}, where $r_t$ is the nanotube radius in meters. A weak RBM was observed in the Raman spectra of this nanotube at 240 cm$^{-1}$, which corresponds to a nanotube diameter of 0.96 nm by the relation $\omega_{RBM}=204/d_t+27$ \cite{meyer_raman_2005}. This yields a thermal conductance of 4 nW/K for the nanotube assuming near equilibrium conditions. From the relation $\stackrel{.}{Q}=G_{th}\Delta T$, the lattice cannot dissipate 6 $\mu$W of power without heating to extremely high temperatures. However, such lattice heating is not supported by our observation, as discussed earlier.

\begin{figure}[b] 
  \centering
  \includegraphics[width=2.5in,keepaspectratio]{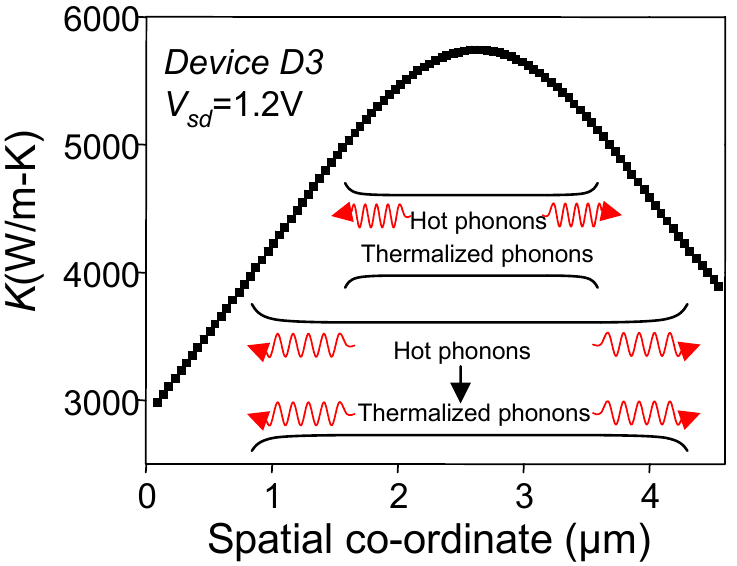}
  \caption{(Color online) Calculated spatial variation of thermal conductivity for device D3. Inset: Proposed mechanism for thermal transport in short (top) and long (bottom) devices.}
  \label{fig:fig4}
\end{figure}

We explain this apparent mismatch between heat generation and dissipation by considering the role of hot phonons. The estimate for $G_{th}$ assumes that all phonon modes are in thermal equilibrium. Using the Landauer model for phonon transport to calculate the thermal power conducted \cite{yamamotoPRL2004},
\begin{eqnarray}
\stackrel {.}{Q}_{ph} = \sum_m\int_0^\infty \frac{dk}{2\pi}\hbar\omega_m(k)v_m(k)\eta(\omega_m,T_{hot})\zeta(\omega_m)
\end{eqnarray}
we can estimate an upper bound for the heat transport by hot modes in the ballistic phonon limit. In equation (2), $\eta$ is the Bose-Einstein distribution function, $\zeta$ is the transmission set to 1 (see \textit{e.g.} ref. \onlinecite{yamamotoPRL2004}), $v_m$ is the group velocity of the $m$th phonon mode and $\hbar\omega_m$ is the phonon energy. Since the effective temperatures under consideration ($T_{hot}\sim$900 K) are much smaller than $\hbar\omega_m$ ($\sim$2000 K), the distribution can be taken to be constant, $\eta\simeq$0.1. Integrating over the range of energies of these phonons (1300-1600 cm$^{-1}$), equation (2) reduces to $\stackrel {.}{Q}_{ph} \simeq 0.4 N\eta$ $\mu$W, where $N$ is the total number of hot phonon modes. Thus, it is possible to dissipate the generated heat if $\simeq$150 phonon modes participate in heat transport. This is smaller than the total number of phonon branches of D2 and hence reasonable. Figure 4(inset) shows a schematic of this heat removal mechanism.

With this understanding of hot-phonon decay and thermal transport for the strongly non-equilibrium case, we now consider longer devices, which show conditions closer to equilibrium. As Fig. 4(inset) shows, the bulk of the lattice becomes heated because the devices are long enough for the hot phonons to thermalize. In this case, the lattice could carry away a significant proportion of the heat generated. Since the lattice is almost as hot as the non-equilibrium phonons, we can estimate the thermal conductivity in the limit where all the heat is dissipated through the lattice. We use an iterative Fourier's law approach along with the Landauer model for electron transport, as developed in ref. \onlinecite{popPRL2005}. Following their formulation:
\begin{equation}
A\frac{d}{dx}(\kappa(x)\frac{dT_{ac}}{dx})+I^2\frac{dR(\lambda_{eff})}{dx}=0
\end{equation}
with
\begin{equation}
\lambda_{eff}(T_{ac}(x),T_{op}(x)) = (\lambda_{ac}^{-1}(x)+\lambda_{op,ems}^{-1}(x)+\lambda_{op,abs}^{-1}(x))^{-1}
\end{equation}
Here, $A$ is the cross-sectional area of the nanotube, $\kappa(x)$ is the thermal conductivity and $R(x)$ is the resistance, corrected for contact resistance $\sim$25 k$\Omega$ and dependent on the effective electron mean-free-path $\lambda_{eff}(x)$. $\lambda_{eff}(x)$ is a function of the lattice temperature $T_{ac}$, hot phonon temperature $T_{op}$ (which are known from our measurement) and scattering lengths $\lambda_{ac,RT}$, $\lambda_{op,min}$ \cite{popPRL2005}. The spatial variation of the parameters necessitates the use of a finite element iterative calculation. Using suitable values for the scattering lengths (as obtained from experiment \cite{parkjyNL2004,javeyPRL2004}), $\lambda_{ac,RT}\sim$1.6 $\mu$m and $\lambda_{op,min}\sim$180 nm, we can obtain a spatial variation of the thermal conductivity. Note that the thermal conductivity as obtained from Fourier's law (eqn. 3) is extremely sensitive to the diameter of the nanotube. While device D1 did not show an RBM in its Raman spectrum, similar data was obtained from device D3 ($L$=4.6 $\mu$m, $\omega_{RBM}$=124 cm$^{-1}$, $d_t$=2.1 nm). The above analysis yields a spatial variation of thermal conductivity for D3 as shown in Fig. 4, roughly following the spatial temperature profile.

We note that the magnitudes of $\kappa$ are similar to other measurements on single-walled nanotubes \cite{yuNL2005}, thus validating the assumption in longer devices that the lattice is the dominant heat carrier. The hot-phonon thermalization length is thus larger than 1 $\mu$m and $\sim$2.3 $\mu$m, considering $L/2$. Note that $\kappa$ shown in Fig. 4 approaches 6000 W/m-K at the maximum (where $T\sim$900 K) and is among the highest reported for any material \cite{footnote_umklapp}.

Our data also allows a direct estimate of thermal contact resistance, $R_{th}$. Previously, laser heating of nanotubes was used to determine of the ratio of thermal contact resistances the left (L) and right (R) lead, $R_{th,L}/R_{th,R}$ \cite{hsuAPL2008}. In our case, since both temperature and heat flow are known, we can directly compute both $R_{th,L}$ and $R_{th,R}$. For D1, we obtain $R_{th,L}\simeq8\times10^7$ K/W and $R_{th,R}\simeq10^7$ K/W respectively. Similar values are obtained for other devices. The asymmetry is device specific and likely dependent on the nanotube-metal contact interface. Thermal contact resistance, thus, accounts for a significant temperature drop at the ends of the nanotube.

In summary, the spatial temperature profile of electrically-heated single-walled nanotubes is obtained for the first time. This measurement provides insights into the mechanism of thermal transport and gives a spatial measure of the thermal conductivity and thermal contact resistances of carbon nanotubes.
\begin{acknowledgments}
We acknowledge useful discussions with James Hone and Philip Kim. Work done at USC was supported by DOE Award No. DE-FG02-07ER46376 and the National Science Foundation Graduate Research Fellowship Program.
\end{acknowledgments}

\end{document}